\documentclass[preprint]{aastex}

\begin{document}
\title{Europium Isotope Ratios in s-Process-Element-Enhanced, Metal-Poor Stars: A
New Probe of the $^{151}$Sm Branching\footnote{Based on data collected
at the Subaru Telescope, which is operated by the National Astronomical
Observatory of Japan.}}

\author{Wako Aoki\altaffilmark{2}, Sean G. Ryan\altaffilmark{3},
Nobuyuki Iwamoto\altaffilmark{2,8}, Timothy
C. Beers\altaffilmark{4},
John E. Norris\altaffilmark{5}, Hiroyasu Ando\altaffilmark{2},
Toshitaka Kajino\altaffilmark{2},
Grant J. Mathews\altaffilmark{6}, 
Masayuki Y. Fujimoto\altaffilmark{7}}

\altaffiltext{2}{National Astronomical Observatory, Mitaka, Tokyo,
181-8588 Japan; email: aoki.wako@nao.ac.jp, ando@optik.mtk.nao.ac.jp,
kajino@nao.ac.jp}
\altaffiltext{3}{Department of Physics and Astronomy, The Open
University, Walton Hall, Milton Keynes, MK7 6AA, UK; email:
s.g.ryan@open.ac.uk} 
\altaffiltext{4}{Department of Physics and Astronomy, Michigan State
University, East Lansing, MI 48824-1116; email: beers@pa.msu.edu}
\altaffiltext{5}{Research School of Astronomy and Astrophysics, The
Australian National University, Mount Stromlo Observatory, Cotter
Road, Weston, ACT 2611, Australia; email: jen@mso.anu.edu.au}
\altaffiltext{6}{Department of Physics, Center for Astrophysics,
University of Notre Dame, Notre Dame, IN 46556, USA; email:
gmathews@nd.edu} 
\altaffiltext{7}{Department of Physics, Hokkaido University, Sapporo,
Hokkaido 060-0810, Japan; email: fujimoto@astro1.sci.hokudai.ac.jp}
\altaffiltext{8}{Present address: Department
of Astronomy, School of Science, University of Tokyo, Hongo,
Bunkyo-ku, Tokyo, 133--0033 Japan, niwamoto@astron.s.u-tokyo.ac.jp}

\begin{abstract} 

We report on the first measurement of the Eu isotope fractions
($^{151}$Eu and $^{153}$Eu) in s-process-element-enhanced, metal-poor
stars. We use these ratios to investigate the $^{151}$Sm branching of
s-process nucleosynthesis. The measurement was made by detailed study
of \ion{Eu}{2} lines that are significantly affected by hyperfine
splitting and isotope shifts in spectra of the carbon-rich, very
metal-poor stars LP~625--44 and CS~31062--050, observed with the
Subaru Telescope High Dispersion Spectrograph.  The $^{151}$Eu
fractions [fr($^{151}$Eu) = $^{151}$Eu/($^{151}$Eu+$^{153}$Eu)]
derived for LP~625--44 and CS~31062--050 are 0.60 and 0.55,
respectively, with uncertainties of about $\pm 0.05$. These values are
higher than found in solar-system material, but agree well with the
predictions of recent s-process models. We derive new constraints on
the temperature and neutron density during the s-process based on
calculations of pulsed s-process models for the $^{151}$Eu fraction.

\end{abstract}
\keywords{nuclear reactions, nucleosynthesis, abundances --- stars: abundances --- stars: AGB and post-AGB --- stars: Population II}


\section{Introduction}\label{sec:intro}

The slow neutron capture process (s-process) is 
responsible for about half of the abundances of elements heavier than
the iron peak in solar-system material. The s-process has
traditionally been described by two approaches: the schematic
classical approach \citep[e.g., ][]{kappeler89}, and modeling of the
s-process in thermally pulsing asymptotic giant branch (AGB) stars
\citep[e.g., ][]{straniero95}. In the classical approach, the neutron
exposure is estimated from the abundance patterns of
pure s-process isotopes in solar-system material, or the elemental
abundance patterns in stars enriched in s-process-elements, while the
temperature and neutron density during the s-process can be estimated
by an analysis of the branchings of the
neutron-capture chains where the rates for beta decay and neutron
capture are comparable.

For $^{151}$Sm, whose half-life is about 90 years, the $\beta$-decay
rate is strongly dependent on temperature, while the neutron capture
rate is not. This makes the $^{151}$Sm branching an excellent
thermometer \citep[e.g., ][]{wisshak95}. Previously, this
branching has been analyzed using the $^{152}$Gd and $^{154}$Gd
isotope ratios in solar-system material, which are believed to be
significantly affected by this branching \citep[e.g., ][]{beer88,
wisshak95}. One difficulty in this approach is that these Gd isotopes
are affected by a small amount of contamination from the p-process,
though they are shielded from the r-process.  Hence, an independent
observational constraint on the branching is desired.

The Eu isotope ratio ($^{151}$Eu/$^{153}$Eu) is one possible
constraint. More than 90~\% of Eu in the solar system is provided by
r-process nucleosynthesis \citep[e.g., ][]{kappeler89, arlandini99,
burris00}, which suggests that it is difficult to constrain the
s-process from measurement of the Eu isotopes in {\it solar-system
material}. However, the Eu isotope ratios can also be measured in
individual stars, as high-resolution spectroscopy can partially
resolve the expected hyperfine splitting and isotope
shifts. \citet{sneden02} and \citet{aoki03a} analyzed the Eu isotopes
for r-process-element-enhanced stars. They showed that the isotope
ratios derived for a total of four of these objects agree well with
that in solar-system material. In this Letter, we apply this analysis
to two s-process-element-enhanced, metal-poor stars, and report the
first results of our analysis of Eu isotopes to analyze the $^{151}$Sm
branching.

\section{Observations}\label{sec:obs}

We selected two subgiants (LP~625--44 and CS~31062--050), which were
found to be enriched in s-process-elements in our previous studies
\citep{aoki00,aoki02b} as most suitable for this investigation. These
stars are both very metal-poor ([Fe/H] = --2.7 and [Fe/H] = --2.4,
respectively\footnote{[A/B] $\equiv \log(N_{\rm A}/N_{\rm B})-
\log(N_{\rm A}/N_{\rm B})_{\odot}$, and $\log \epsilon_{\rm A} \equiv
\log(N_{\rm A}/N_{\rm H})+12$ for elements A and B.}) and have quite
similar atmospheric parameters, as shown in Table~\ref{tab:obs}.  The
observed enhancement of s-process-elements in these stars is believed
to be due to nucleosynthesis in extinct AGB stars which transfered
mass to the surviving luminous companions of the binary systems. The
binarity of these stars is discussed below. The enhancement of the
bulk of the s-process elements in these two stars is also similar
(e.g., [Ba/Fe]$\sim +2.5$). One important difference between these two
objects is their Pb abundance: the Pb/Ba ratio in CS~31062--050 is
five times higher than that in LP~625--44. For comparison purposes, we
also selected a very metal-deficient giant, HD~6268 ([Fe/H] = --2.5),
which exhibits a moderate enhancement of r-process-elements, and has
strong Eu lines, comparable to those in the two
s-process-element-enhanced stars.

The [Ba/Eu] values of LP~625--44 and CS~31062--050 derived in our
previous studies \citep{aoki02a,aoki02b} are 1.09 and 0.46,
respectively. These are more than 1~dex higher than the value of the
r-process component in solar-system material ($-0.69$, Arlandini et
al. 1999). While the [Ba/Eu] value of LP~625--44 is similar to that of
the solar-system s-process component ($+1.15$), the value of
CS~31062--050 is significantly lower. However, low [Ba/Eu] values
($\sim 0.4$) are predicted by some AGB nucleosynthesis models
\citep[e.g., ][]{goriely00} for metal-deficient stars, for which high
neutron-exposure is expected due to the high ratios of neutrons per
seed nuclei. Indeed, the value of [Ba/Eu]$\sim 0.4$ is not unusual in
carbon-rich, metal-deficient stars like CS~31062--050 \citep[e.g.,
][]{aoki02b,johnson02}. To explain these [Ba/Eu] values by
contamination from the r-process, we must assume a very large
enhancement of r-process elements ([Eu/Fe]$\sim$ +1.5--1.8), similar
to that in extremely r-process-enhanced stars, such as CS~31082--001
\citep{cayrel01}, which are known to be quite rare.  For these
reasons, we assume in the following discussion that the majority of
the Eu in CS~31062--050, as well as in LP~625--44, originates from the
s-process.

The observations we report here were made with the Subaru Telescope
High Dispersion Spectrograph (HDS; Noguchi et al. 2002) in August
2002. The wavelength range covered was 3100-4700~{\AA}, with a
resolving power $R = 90,000$. The total exposure times and
signal-to-noise ratios obtained are listed in Table~\ref{tab:obs}.

The heliocentric radial velocities measured for our spectra are given
in Table~\ref{tab:obs}. \citet{aoki00} already reported a variation of
radial velocities for LP~625--44, which is confirmed by the present
measurement. For CS~31062-050, a change of radial velocity from the
previous measurement ($V_{r} = 7.75$~km~s$^{-1}$ by Aoki et al. 2002b)
was also detected. These results suggest the binarity of these stars,
and strongly support the mass-transfer scenario for the enrichment of
s-process elements.

\section{Analysis and Results}\label{sec:ana}

We adopted the Eu line data, including the hyperfine splitting and
isotope shifts, provided by \citet{lawler01}. The isotope fractions of
$^{151}$Eu [fr($^{151}$Eu) = $^{151}$Eu / ($^{151}$Eu + $^{153}$Eu)]
were measured by fitting observed spectra with synthetic ones
calculated for the Eu lines, using the model atmospheres of
\citet{kurucz93} and atmospheric parameters derived in the previous
studies \citep{aoki02a,aoki02b}. The macro-turbulent velocity was
estimated by fitting a gaussian profile to clean Fe and Ti lines
detected in the same spectrum for each object (see
Table~\ref{tab:obs}). Our careful investigation of the HDS
instrumental profiles around the Eu lines used in the analysis showed
that the profile is quite symmetric. Hence, a gaussian profile
approximation is sufficient for the analysis of stellar spectra that
are further broadened by macro-turbulence.

\begin{figure} 
\includegraphics[width=8cm]{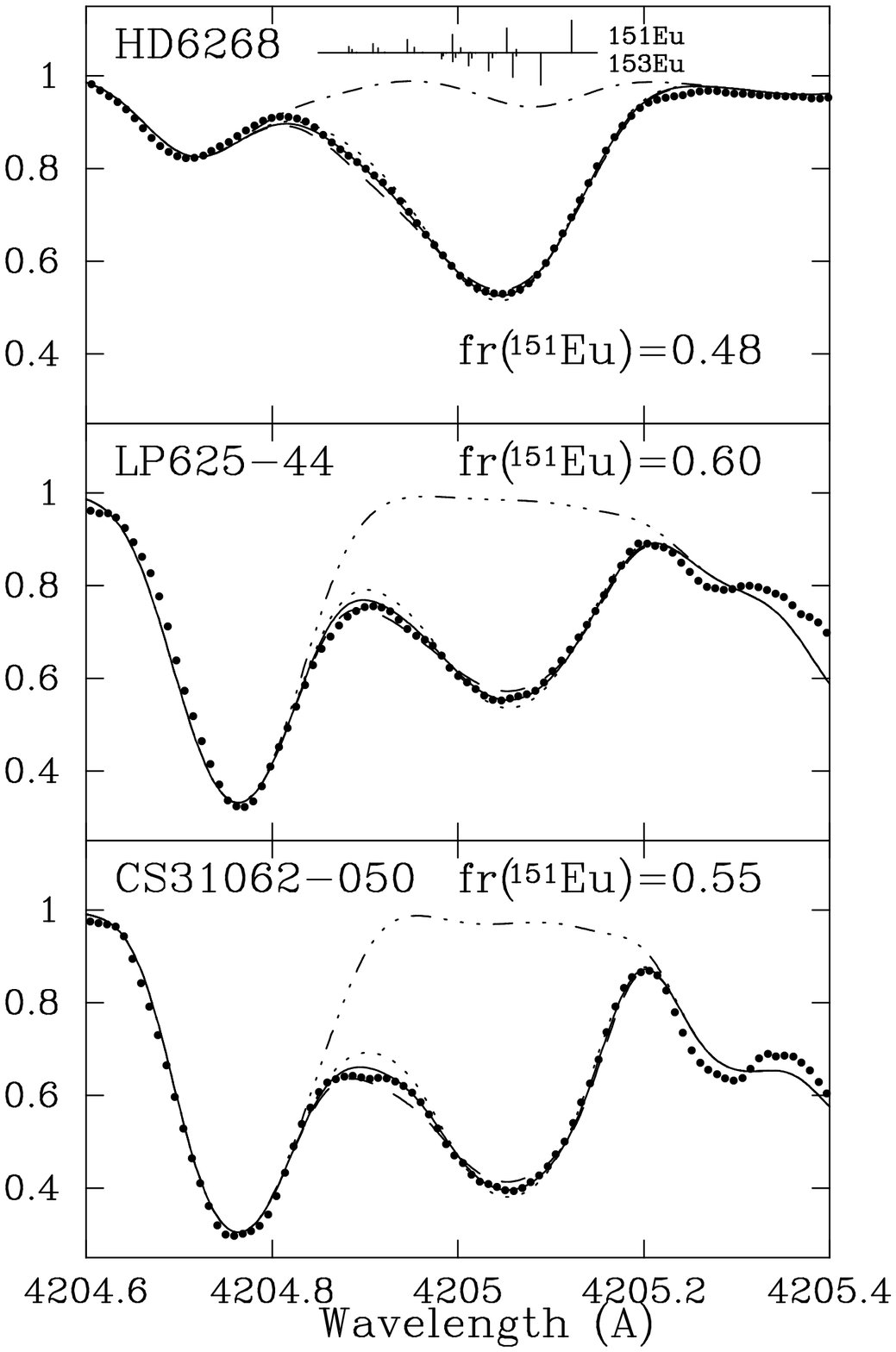} 
\caption[]{Comparison of the observed spectra (dots) and synthetic
ones (lines) for the \ion{Eu}{2} 4205~{\AA} line. The name of the
object and the adopted fr($^{151}$Eu) value are presented in each
panel. The solid line shows the synthetic spectra for the adopted
fr($^{151}$Eu); the dotted and dashed lines show those for ratios
which are smaller and larger by 0.10 in fr($^{151}$Eu),
respectively. The dot-dashed lines show the synthetic spectra for no
Eu. The weak blending at 4205.09~{\AA} in the spectrum of HD~6268
(top) is due to \ion{V}{2}, whose effect is negligible in the two
subgiants.. The wavelengths and relative strength of the hyperfine
components for $^{151}$Eu and $^{153}$Eu are shown in the top panel.}
\label{fig:sp} 
\end{figure}

Figure~\ref{fig:sp} shows the spectra of the \ion{Eu}{2} 4205~{\AA}
line in our three program stars. In the calculation of synthetic
spectra, blending by other species is included using the comprehensive
line list of \citet{kurucz95} and the CH and CN line lists produced in
\citet{aoki02b}.  The dot-dashed lines show the synthetic spectra
calculated without the \ion{Eu}{2} contribution. The absorption around
4205~{\AA} is dominated by \ion{Eu}{2}. In the top panel, the line
positions of both Eu isotopes are shown. Since the hyperfine splitting
of $^{151}$Eu is larger than that of $^{153}$Eu, the asymmetry of the
line profile increases with increasing fr($^{151}$Eu).

Our measurements were carried out using the same procedure as was
utilized for r-process-enhanced stars in \citet{aoki03a}. We searched
for the isotope fractions that minimize the value of $\chi_{\rm
r}^{2}$ for a given Eu abundance. We estimated the uncertainty of the
derived Eu isotope fractions and total abundances by considering the
range over which the $\chi_{\rm r}^{2}$ is twice as large as the
best-fit case\footnote{This corresponds to approximately a $3 \sigma$
confidence level for 25 degrees of freedom, which is the case of our
analysis for Eu lines. We do not, however, insist on the exact value
of the confidence level, because the statistical properties are not
proven for spectra re-binned to a constant wavelength step (Bonifacio
\& Caffau 2003).}.  The error in the isotope fraction due to the
uncertainty in the total Eu abundance was estimated from the range of
the isotope fraction allowed within the adopted abundance uncertainty.

We estimated the errors arising from the following factors using the
same procedures as in \citet{aoki03a}: (1) The macro-turbulent
velocity (given in Table~\ref{tab:obs}); (2) The continuum-level
uncertainty estimates, which are assumed to be 0.5\% for HD~6268 and
1\% for the other two stars; and (3) The wavelength calibration of the
spectrum and the Eu line position. The wavelength shift was estimated
from the longer wavelength (redder) part of the spectral line, which
is insensitive to the Eu isotope ratio; the error is typically a few
m{\AA}.

We analyzed three \ion{Eu}{2} lines at 3819, 4129, and 4205~{\AA}.
Table~\ref{tab:res} gives the derived $^{151}$Eu fraction and total
error ($\sigma_{\rm total}$), estimated from the quadrature sum of the
individual errors mentioned above. The Eu ($^{151}$Eu + $^{153}$Eu)
abundance derived from each line is also given in the table. The
4205~{\AA} line is most sensitive to the Eu isotope
fractions. Therefore, the fractions deduced from this line have the
smallest uncertainty, even though the strong CH line at 4204.75~{\AA}
affects the bluest part of the \ion{Eu}{2} line. The 3819~{\AA} line
is strongest among the three lines. However, the uncertainty in the
derived isotope ratio due to the error in the Eu abundance estimation
is significant \citep{aoki03a}. The \ion{Eu}{2} 4129~{\AA} line, as
for \ion{Eu}{2} 4205~{\AA}, has suitable strength for this
analysis. However, our synthetic spectra fail to well reproduce the
blue portion ($\sim$ 4129.6~{\AA}) of the absorption line in the two
carbon-rich stars. There seems to exist some unidentified absorption
lines. Excluding the blue wing from the fitting, we derived the
isotope fractions, given in Table~\ref{tab:res}, for the range of
4129.65--4129.82~{\AA}. Though the number of data points within this
range is sufficient, the results are rather sensitive to the choice of
wavelength range used for the analysis.

For these reasons, we prefer to adopt the isotope fractions derived
from the \ion{Eu}{2} 4205~{\AA} line as the best determination. We
note, for comparison purposes, that the weighted means of the results
from all three lines for HD~6268, LP~625--44, and CS~31062--050 are
0.47, 0.61, and 0.57, respectively, which agree well with the results
from the 4205~{\AA} line alone.

The fr($^{151}$Eu) of HD~6268 (0.48$\pm 0.04$) perfectly agrees with
that of solar-system material \citep[0.478; ][]{anders89}, as found in
other r-process-element-enhanced stars \citep{sneden02,aoki03a}. In
contrast, the fr($^{151}$Eu) of the two stars enriched in s-process
elements are higher than the solar-system value.

\section{Discussion}

The yields of the s-process nuclides which explain the abundances of
pure s-process isotopes in solar-system material have been calculated
by \citet{arlandini99}. The fr($^{151}$Eu) values deduced from their
best-fit stellar and classical models are 0.541 and 0.585,
respectively. The first conclusion of the present investigation is
that {\it the agreement between these values and those derived from
our observations is quite good}. 

In order to further investigate what can be learned about the
s-process from the measured Eu isotopes, we have made an analysis
using the thermally pulsed s-process models described in Howard et al.
(1986). We utilized updated neutron-capture rates \citep[e.g.,
][]{bao00}. In particular, in order to follow the nuclear flow in the
Sm-Eu-Gd region, we included the rates for $^{151-155}$Eu and
$^{151}$Sm obtained by \citet{best01} and \citet{toukan95}. The
electron-capture rate of $^{153}$Gd was adopted from
\citet{takahashi87}. We found that the fr($^{151}$Eu) value has almost
no dependence on the mean neutron exposure ($\tau_{0}$) for a fixed
temperature in the range of $0.2\leq \tau_{0} \leq 0.8$, which well
covers the $\tau_{0}$ ($\sim 0.5$~mb$^{-1}$) estimated by
\citet{aoki01} for LP~625--44. Our results are calculated with
$\tau_{0}=$ 0.3~mb$^{-1}$, which gives a good fit to solar abundances.

\begin{figure}
\includegraphics[width=8cm]{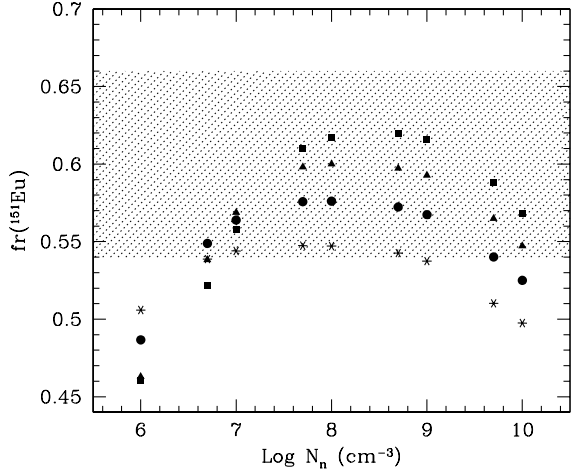}
\caption[]
{The fr($^{151}$Eu) values calculated for four temperatures (squares:
$kT$ = 30~keV, triangles: 20~keV, circles: 15~keV, and asterisks:
10~keV) as a function of neutron density ($N_{\rm n}$). The
fr($^{151}$Eu) of LP~625--44 and its uncertainty is shown by the
hatched area. } 
\label{fig:fr151eu} 
\end{figure}

Figure~\ref{fig:fr151eu} shows the fr($^{151}$Eu) values calculated by
our model. They are plotted as a function of neutron density ($N_{\rm
n}$) for four temperatures ($kT=$ 10, 15, 20, and 30~keV).  Also
shown for comparison by the hatched area is the fr($^{151}$Eu) range
deduced for the s-process-element-enhanced star LP~625--44. As can be
seen in this figure, the fr($^{151}$Eu) value is rather sensitive to
the ambient temperature and neutron density during the s-process. The
fr($^{151}$Eu) value is maximized in the range of neutron density from
$N_{\rm n} =$ 5$\times$ $10^7$ to $10^9$~cm$^{-3}$.

The branching factor at $^{151}$Sm is given by
$f=\lambda_n/(\lambda_\beta+\lambda_n)$ , where $\lambda_n = N_{\rm n}
v_{T} <\sigma>$ and $\lambda_\beta = \ln 2/t_{1/2}$ are the
neutron-capture rate and the $\beta$-decay rate, respectively. In
these formulations, $v_{T}$, $<\sigma>$, and $t_{1/2}$ are the mean
thermal velocity, the Maxwellian averaged cross section, and the
half-life, respectively. For $N_{\rm n} > 10^{7}$~cm$^{-3}$, the branching
factor is higher than 0.9. This indicates that the nuclear flow
bypasses $^{151}$Eu. In this case, the $^{151}$Eu abundance is
produced by the decay of $^{151}$Sm after the neutron-capture flow
ceases. As the neutron density increases to $N_{\rm n} > 10^{9}$~cm$^{-3}$,
another branching at $^{153}$Sm becomes important. Since we adopt a
much smaller cross section for $^{153}$Sm than those for $^{151}$Sm
and $^{153}$Eu, the $^{153}$Sm abundance relative to $^{151}$Sm and
$^{153}$Eu {\it during} the neutron capture reactions increases with
increasing neutron density. This results in a decrease of the {\it
final} value of fr($^{151}$Eu) with increasing neutron density for
$N_{\rm n} > 10^{9}$~cm$^{-3}$, as found in Figure~\ref{fig:fr151eu}. A
measurement of the cross section for $^{153}$Sm is thus highly
desirable to constrain the high neutron density branching.

In contrast to the high neutron-density conditions, $^{151}$Sm
$\beta$-decay {\it during} the s-process is increasingly important
with decreasing neutron density ($N_{\rm n}\lesssim 10^{7}$
~cm$^{-3}$). For the typical temperature ranges thought to apply in
the s-process (10-30~keV), the neutron-capture rate on $^{151}$Eu is
faster than that for $^{153}$Eu. Therefore, the nuclear flow passes
through $^{151}$Eu rather easily, and creates $^{153}$Gd via
$^{151}$Eu($n$, $\gamma$) $^{152}$Eu($\beta,\nu$)$^{152}$Gd($n$,
$\gamma$)$^{153}$Gd.  In low neutron-density conditions, the
electron-capture on $^{153}$Gd is comparable with, or faster than, the
neutron capture. This contributes to the production of $^{153}$Eu and
explains the decrease of fr($^{151}$Eu) with decreasing neutron
density in the low ($N_{\rm n}\lesssim 10^{7}$ cm$^{-3}$) range. This trend
appears more clearly for the higher temperature case, because the
effect of a higher neutron-capture rate on $^{151}$Eu relative to that
on $^{153}$Eu is large, while the branching factor decreases with
increasing temperature.

The comparison of the calculations with the observational results
indicates that s-processes with high neutron density ($\log N_{\rm n}
\gtrsim 9$) and low temperature ($kT \lesssim 20$~keV), or those with
quite low neutron density ($N_{\rm n}<10^{7}$cm~s$^{-1}$), are not
allowed. This is a new constraint on s-process nucleosynthesis
provided by the Eu isotope analysis.

Recent models of AGB stars show that the abundance patterns of nuclei
in branchings are affected by the s-process both during the thermal
pulses and between pulses \citep[e.g., ][]{arlandini99}. The reaction
which provides neutron in the former phase is
$^{22}$Ne$(\alpha,n)^{25}$Mg, which produces a neutron density as high
as $10^{8}$--$10^{10}$~cm$^{-3}$. In the interpulse phase,
$^{13}$C$(\alpha,n)^{16}$O is assumed to be the neutron source
reaction, which leads to a lower neutron density ($N_{\rm n}\sim
10^{7}$~cm$^{-3}$). The observational errors of the derived Eu isotope
fractions are still too large to constrain the contribution of each
process to the final abundance patterns produced by nucleosynthesis in
AGB stars. However, our analysis shows that Eu isotopes can be a new
probe to determine the temperature and neutron density in s-process
nucleosynthesis.

We note that the fr($^{151}$Eu) values derived by using the previous
nuclear data of Sm and Eu isotopes, in particular the $^{151}$Sm
neutron-capture rate, are significantly lower than the results shown
above. The previous neutron-capture rate recommended by \citet{bao00},
e.g., 2377~mb at 30~keV, is significantly higher than the rate used in
the present work (1585~mb at 30~keV, Best et al. 2001). The high
neutron-capture rate causes the nuclear flow to bypass $^{151}$Eu, and
results in lower fr($^{151}$Eu) values by 0.05-0.08. The low
neutron-capture rate is clearly preferable to explain our
observational results. New experiments to determine the neutron
capture cross section of $^{151}$Sm are highly desirable to fix ratios
of $^{151}$Eu/$^{153}$Eu and also $^{152}$Gd/$^{154}$Gd.

It should also be noted that the Eu isotope fractions are quite
similar between LP~625--44 and CS~31062--050, even though their Pb/Ba
abundance ratios are significantly different \citep{aoki02a,aoki02b}.
This observational fact indicates that nuclear flow is very fast, and
fr($^{151}$Eu) easily obtains asymptotic values in the Sm-Eu-Gd
region, once the flow passes through the $N=82$ neutron magic
nuclei. This situation is quite likely to occur in these very
metal-poor stars because of the expected high neutron-to-seed
ratio. However, the Pb abundance is very sensitive to the neutron
exposure \citep[e.g., ][]{gallino03}, and large variations of the
Pb/Ba ratio are expected. Similar studies of additional
s-process-enhanced stars would be useful to derive more clear
conclusions on production of heavy nuclei by the s-process.

\acknowledgments

%

%

\begin{deluxetable}{@{}l@{\extracolsep{\fill}}c@{\extracolsep{\fill}}c@{\extracolsep{\fill}}c@{\extracolsep{\fill}}c@{\extracolsep{\fill}}c@{\extracolsep{\fill}}c}
\tabletypesize{\small}
\tablecaption{PROGRAM STARS AND OBSERVATIONS\label{tab:obs}}
\startdata
\tableline
\tableline
Star          &  Exp.\tablenotemark{a}  & S/N\tablenotemark{b} &  Obs. date (JD) & Radial velocity ~ & $T_{\rm eff}/\log g/$[Fe/H]/$v_{\rm micro}$ & $v_{\rm macro}$ \\ 
              &          &     &                         & (km s$^{-1}$)    &                  & (km s$^{-1}$)   \\
\tableline
LP~625--44    & 270 (9)  & 137 & 22/08/2002 (2,452,509) & $28.41\pm0.19$ & 5500/2.5/$-2.7$/1.2 & $6.69\pm 0.13$ \\
CS~31062--050 ~ & 390 (13) & 124 & 23/08/2002 (2,452,510) & $11.51\pm0.17$ & 5600/3.0/$-2.4$/1.3 & $5.30\pm 0.40$ \\
HD~6268       & 100 (5)  & 304 & 22/08/2002 (2,452,509) & $40.13\pm0.15$ & 4600/1.0/$-2.5$/2.1 & $7.27\pm 0.27$ \\
\tableline
\enddata

\tablenotetext{a}{
Total exposure time (minute) and number of exposures.
}

\tablenotetext{b}{
S/N ratio per 0.012~{\AA} pixel at 4000~{\AA}. 
}

\end{deluxetable}

\begin{deluxetable}{ccccccccc}
\tablecaption{RESULTS\label{tab:res}}
\startdata
\tableline
\tableline
 line  &  \multicolumn{2}{c}{HD~6268} & & \multicolumn{2}{c}{LP~625--44} & & \multicolumn{2}{c}{CS~31062--050} \\
\cline{2-3}\cline{5-6}\cline{8-9}
       & fr($^{151}$Eu) & $\log \epsilon$(Eu) & & fr($^{151}$Eu) & $\log \epsilon$(Eu) & & fr($^{151}$Eu) & $\log \epsilon$(Eu) \\
\tableline
3819{\AA} & 0.44 $\pm$ 0.11 & $-1.55$ & & 0.68 $\pm$ 0.10 & $-0.44$ & & 0.65 $\pm$ 0.14 & $-0.02$ \\
4129{\AA} & 0.49 $\pm$ 0.07 & $-1.57$ & & 0.56 $\pm$ 0.07 & $-0.45$ & & 0.56 $\pm$ 0.09 & $+0.01$ \\
4205{\AA} & 0.48 $\pm$ 0.04 & $-1.54$ & & 0.60 $\pm$ 0.06 & $-0.43$ & & 0.55 $\pm$ 0.05 & $+0.04$ \\
\tableline
\enddata
\end{deluxetable}

\end{document}